\documentclass[aps,twocolumn,preprintnumbers,amsmath,amssymb,superscriptaddress]{revtex4-2}

\usepackage{graphicx}
\usepackage{dcolumn}
\usepackage[colorlinks=true, linkcolor=blue, citecolor=blue, urlcolor=blue]{hyperref}

\usepackage{amsmath}
\usepackage{xcolor}
\usepackage{soul}
\usepackage{graphicx}
\usepackage{dcolumn}
\usepackage{bm}
\def\lesssim{\ \raise.3ex\hbox{$<$}\kern-0.8em\lower.7ex\hbox{$\sim$}\ }
\def\gesim{\ \raise.3ex\hbox{$>$}\kern-0.8em\lower.7ex\hbox{$\sim$}\ }

\begin{document}
\preprint{RIKEN-iTHEMS-Report-23}

\title{Non-Hermitian $p$-wave superfluid and effects of the inelastic three-body loss in a one-dimensional spin-polarized Fermi gas}
\author{Hiroyuki Tajima}
\affiliation{Department of Physics, Graduate School of Science, The University of Tokyo,
    Tokyo 113-0033, Japan}
\author{Yuta Sekino}
\affiliation{Interdisciplinary Theoretical and Mathematical Sciences Program (iTHEMS), RIKEN, Wako, Saitama 351-0198, Japan}
\affiliation{Nonequilibrium Quantum Statistical Mechanics RIKEN Hakubi Research Team, RIKEN Cluster for Pioneering Research (CPR), Wako, Saitama 351-0198, Japan}
\affiliation{RIKEN Cluster for Pioneering Research (CPR), Astrophysical Big Bang Laboratory (ABBL), Wako, Saitama, 351-0198 Japan}
\author{Daisuke Inotani}
\affiliation{Departments of Physics, Keio University, 4-1-1 Hiyoshi, Kanagawa 223-8521, Japan}
\author{Akira Dohi}
\affiliation{RIKEN Cluster for Pioneering Research (CPR), Astrophysical Big Bang Laboratory (ABBL), Wako, Saitama, 351-0198 Japan}
\affiliation{Interdisciplinary Theoretical and Mathematical Sciences Program (iTHEMS), RIKEN, Wako, Saitama 351-0198, Japan}
\author{Shigehiro Nagataki}
\affiliation{RIKEN Cluster for Pioneering Research (CPR), Astrophysical Big Bang Laboratory (ABBL), Wako, Saitama, 351-0198 Japan}
\affiliation{Interdisciplinary Theoretical and Mathematical Sciences Program (iTHEMS), RIKEN, Wako, Saitama 351-0198, Japan}
\affiliation{Astrophysical Big Bang Group (ABBG), Okinawa Institute of Science and Technology Graduate University (OIST), Tancha, Onna-son, Kunigami-gun, Okinawa 904-0495, Japan}
\author{Tomoya Hayata}
\affiliation{Departments of Physics, Keio University, 4-1-1 Hiyoshi, Kanagawa 223-8521, Japan}
\affiliation{Interdisciplinary Theoretical and Mathematical Sciences Program (iTHEMS), RIKEN, Wako, Saitama 351-0198, Japan}
\date{\today}
\begin{abstract}
We theoretically investigate non-Hermitian $p$-wave Fermi superfluidity in one-dimensional spin-polarized Fermi gases which is relevant to recent ultracold atomic experiments.
Considering an imaginary atom-dimer coupling responsible for the three-body recombination process in the Lindblad formalism,
we discuss the stability of the superfluid state against the atomic loss effect. 
Within the two-channel non-Hermitian BCS-Leggett theory, the atomic loss is characterized by the product of the imaginary atom-dimer coupling and the $p$-wave effective range.
Our results indicate that for a given imaginary atom-dimer coupling,
a smaller magnitude of the effective ranges of $p$-wave interaction is crucial for reaching the non-Hermitian $p$-wave Fermi superfluid state.  
\end{abstract}

\maketitle


\section{Introduction}
\label{sec:1}
Unconventional superconductivity and superfluidity are of great interest due to their non-trivial properties such as anisotropic gap structures and topological states. 
In particular, $p$-wave superconductors and superfluids have been discussed in the various contexts of physics ranging from condensed-matter physics\cite{RevModPhys.63.239} to nuclear physics~\cite{RevModPhys.75.607}.
Once stable topological $p$-wave superconductors or superfluids can be manipulated in a controllable manner,
it would make significant progress toward the realization of a universal quantum computer based on anyons~\cite{beenakker2016road}.

In this regard, over the last few decades, great efforts have been devoted to experimentally realizing the $p$-wave superfluid state in ultracold atoms~\cite{PhysRevLett.90.053201,PhysRevA.70.030702,PhysRevA.77.053616,PhysRevLett.101.100401,PhysRevLett.120.133401,PhysRevA.98.020702,PhysRevA.99.052704,PhysRevA.104.043311,PhysRevA.106.023322,PhysRevLett.125.263402,marcum2020suppression,PhysRevA.107.053310,venu2023unitary}, because of the strong advantages of the atomic systems with tunable pairing interaction near the Feshbach resonance~\cite{RevModPhys.82.1225}.
Moreover, the transition from the molecular Bose-Einstein condensates (BEC) to Bardeen-Cooper-Schrieffer (BCS) Fermi superfluid in the $p$-wave channel has been discussed  theoretically~\cite{PhysRevLett.94.230403,PhysRevLett.94.050403,PhysRevLett.96.040402,PhysRevLett.99.210402,PhysRevA.85.053628,PhysRevA.92.063638}.

However,
the experimental realization of $p$-wave Fermi superfluids in ultracold atoms has not been achieved yet.
One of the major obstacles is the inevitable three-body loss induced by the three-body recombination process where two atoms form a deep bound state
and its binding energy is released by another energetic atom~\cite{PhysRevLett.90.053201}.
For the narrow resonance, the atomic loss rate is well explained by the so-called two-step cascade model~\cite{PhysRevLett.120.193402,PhysRevA.99.052704,PhysRevLett.125.263402}.
Recently, one-dimensional systems have attracted attention since it is theoretically reported that the atomic loss associated with the $p$-wave interaction can be suprressed~\cite{PhysRevA.95.032710,PhysRevA.96.030701,PhysRevA.98.011603,PhysRevA.102.043319,PhysRevA.106.043310}.
While the suppression of atomic losses in one-dimensional systems is under investigation in recent experiments~\cite{PhysRevLett.125.263402,marcum2020suppression},
it is worth investigating theoretically how close the present experimental situation is to realizing the $p$-wave superfluid state in one dimension~\cite{PhysRevA.102.013307,PhysRevB.105.064508}.

To explore the effects of particle losses on the many-body supefluid state, the non-Hermitian BCS formalism can be a promising route~\cite{PhysRevLett.123.123601}.
By using the complex-valued interaction,
one can incorporate the effect of the atomic loss in the BCS formalism in a similar manner to conventional Hermitian models. 
While the quantum jump term is neglected in the non-Hermitian BCS theory,
this treatment can be justified in postselected quantum trajectories~\cite{daley2014quantum} by separating the transient dynamics into the non-unitary evolution and the quantum jump process, where one may selectively observe ensembles remaining in the system without experiencing the jump process.

The BCS theory is further extended to describe the BCS-BEC crossover by solving the particle-number equation with respect to the chemical potential self-consistently, which is referred to as the BCS-Leggett theory~\cite{ohashi2020bcs}.
The non-Hermitian extension of the BCS-Leggett theory, which covers not only the weak-coupling BCS regime but also the strong-coupling BEC regime in the continuum model~\cite{PhysRevA.103.013724,PhysRevA.107.033331}, can be useful in studying the impact of particle losses in the present system.
The present authors applied this approach to one-dimensional two-component $p$-wave Fermi superfluid with the dominant two-body loss originating from the dipolar relaxation~\cite{PhysRevA.107.033331}.

However, to our knowledge, there are no previous studies on non-Hermitian many-body states with the three-body loss in spin-polarized Fermi gases even though the three-body loss is a major obstacle in ultracold atomic systems.
Obviously, the three-body loss effect is theoretically challenging since it may involve a kind of effective three-body interaction~\cite{sekino2018comparative,PhysRevA.102.053304,PhysRevA.103.L021302,PhysRevA.103.043307,PhysRevA.108.043303,wang2023three} which is not well established compared with the two-body interaction in the present system.

In this work,
we theoretically discuss the stability of the $p$-wave superfluid state against the three-body loss in a one-dimensional spin-polarized Fermi gas by using the non-Hermitian BCS-Leggett theory. 
The imaginary atom-dimer coupling can describe the inelastic atom-dimer collision responsible for the three-body recombination.
To incorporate such a non-Hermitian term,
we start from the two-channel Hamiltonian~\cite{gurarie2007resonantly} for the $p$-wave Feshbach resonance where open- and closed-channel atoms are simultaneously considered.
We show that this non-Hermitian two-channel model gives a microscopic description of the two-step cascade process~\cite{PhysRevLett.120.193402,PhysRevA.99.052704,PhysRevLett.125.263402}.
In this framework,
we investigate the possible ground-state phase diagram to see the condition realizing the non-Hermitian $p$-wave Fermi superfluid state under the three-body loss process.

Our study would be useful for the future experimental realization of the $p$-wave Fermi superfluid under the inevitable three-body loss effect in ultracold atoms.
Moreover, our theoretical development of the non-Hermitian many-body theory for open quantum environments with the inelastic three-body loss would make an important step in studying various fascinating topics in the field of ultracold atomic physics.

This paper is organized as follows.
In Sec.~\ref{sec:2}, we first introduce the effective Hamiltonian.
Second, we apply the non-Hermitian BCS-Leggett theory to this model.
Finally, we show how the Lindblad master equation is related to the observed loss.
In Sec.~\ref{sec:3}, we show the numerical results of the ground-state phase diagram of the mean-field theory.
The summary of this paper is given in Sec.~\ref{sec:4}.
Hereafter, we work in the unit of $k_{\rm B}=\hbar=1$ for convenience.

\section{Formalism}
\label{sec:2}
\subsection{Effective Hamiltonian}
We consider the two-channel effective Hamiltonian of one-dimensional homogeneous spinless fermions near the $p$-wave Feshbach resonance with the three-body force, which is given explicitly by~\cite{PhysRevA.101.062702}
\begin{align}
    H_{\rm eff}=H_{\rm a}+H_{\rm d}+V_{2}+V_{3},
\end{align}
where
\begin{align}
    H_{\rm a}=\sum_{k}\xi_{k}a_{k}^\dag a_{k}, \quad 
    H_{\rm d}=\sum_{P}
    \xi_{P,{\rm d}}d_{P}^\dag d_{P},
\end{align}
are kinetic terms of Fermi atoms and Feshbach dimer, respectively.
$\xi_{k}=k^2/(2m)-\mu$ is a single-particle energy of a Fermi atom with a mass $m$ measured from the chemical potential $\mu$ and $\xi_{P,{\rm d}}=P^2/(4m)+\nu-2\mu$ is that of Feshbach dimer with an energy level $\nu$.
$a_{k}^{(\dag)}$ and $d_P^{(\dag)}$ denote annihilation (creation) operators of a Fermi atom and a Feshbach dimer with momenta $k$ and $P$, respectively.

The interaction terms consist of a two-body term for the $p$-wave Feshbach resonance
\begin{align}
    V_2 =\sum_{P,k}
    {g_{2}k}
    d_{P}^\dag a_{-k+P/2}a_{k+P/2}
    +{\rm h.c.\;},
\end{align}
and a non-Hermitian three-body term for the inelastic atom-dimer collision
\begin{align}
    V_3=
    \sum_{Q,q,q'}
    g_{3}
    d_{q+Q/2}^\dag
    a_{-q+Q/2}^\dag
    a_{-q'+Q/2}
    d_{q'+Q/2}.
\end{align}
The two-body coupling constant $g_2$ is related to the $p$-wave scattering length $a$ and effective range $r$ (where the one-dimensional $p$-wave phase shift is defined by $k{\rm cot}\delta(k)=-\frac{1}{a}+rk^2/2$)
through~\cite{PhysRevA.94.063650}
\begin{align}
    \frac{m}{2a}=-\frac{\nu}{2g_2^2}+\frac{m\Lambda}{\pi},
\end{align}
\begin{align}
    r=-\frac{2}{m^2g_2^2},
\end{align}
where $\Lambda$ is the momentum cutoff.
On the other hand, $g_3$ is taken to be a pure imaginary value in this work since we are interested in the effect of inelastic atom-dimer scattering.
Later we relate $g_3$ to the inelastic atom-dimer collision rate $K_{\rm ad}$ which is determined experimentally~\cite{PhysRevLett.125.263402}.
For simplicity, we ignore the off-resonant background interactions.

\subsection{Non-Hermitian BCS-Leggett theory}
In the following, we apply the non-Hermitian BCS-Leggett theory~\cite{PhysRevA.103.013724,PhysRevA.107.033331} to $H_{\rm eff}$ by assuming the postselected quantum trajectories~\cite{daley2014quantum}.
We replace $d_P$ and $d_P^\dag$ in $H_{\rm eff}$ with
a pair of complex mean-field order parameters $\phi$ and $\bar{\phi}$, respectively (note that $\bar{\phi}\neq \phi^*$ due to the non-Hermicity of $H_{\rm eff}$)
and neglect the contributions associated with nonzero center-of-mass momentum of a dimer (i.e., $d_{P\neq 0}$ and $d_{P\neq 0}^\dag$).
Accordingly,
we
obtain the mean-field Hamiltonian
\begin{align}
 H_{\rm MF}&=(\nu-2\mu)\bar{\phi}\phi
    +
    \frac{1}{2}\sum_{k}\tilde{\xi}_{k}\cr
    &
    +\frac{1}{2}
    \sum_{k}\Psi_{k}^\dag
    \left(\begin{array}{cc}
       \tilde{\xi}_{k}  & -\bar{\Delta}_{k} \\
        -\Delta_{k} & -\tilde{\xi}_{k}
    \end{array}\right)
    \Psi_{k},
\end{align}    
where we introduced the shifted dispersion
    $\tilde{\xi}_{\bm{k}}
    =k^2/(2m)-\mu_{\rm eff}$ with the effective chemical potential $\mu_{\rm eff}=\mu-g_3\bar{\phi}\phi\equiv \mu_{\rm eff,R}+i\mu_{\rm eff,I}$,
    the complex pairing gaps $\Delta_k=-2g_2\phi k$ and $\bar{\Delta}_k=-2g_2\bar{\phi}k$, and the Nambu spinor $\Psi_k=\left(a_k \ a_{-k}^\dag\right)^{\rm t}$.
For convenience, we introduce $\phi_0=\phi_{\rm R}+i\phi_{\rm I}$ satisfying $\phi_0^2=\bar{\phi}\phi$ and $D_0=-2g_2\phi_0\equiv D_{\rm R}+iD_{\rm I}$.
Following the previous works~\cite{PhysRevA.103.013724,PhysRevA.107.033331},
we assume $D_{\rm R}\geq 0$, $D_{\rm I}\geq 0$, and $\mu_{\rm eff, I}\leq 0$. 
At $\mu_{\rm eff, R}>0$, the present system encounters the exceptional point when the indicator
\begin{align}
    z=D_{\rm R}\sqrt{m^2D_{\rm I}^2+2m\mu_{\rm eff,R}}-mD_{\rm R}D_{\rm I}+\mu_{\rm eff,I}
\end{align}
becomes zero~\cite{PhysRevA.103.013724,PhysRevA.107.033331}, and the superfluid solution vanishes there.
In turn, at $\mu_{\rm eff, R}<0$,
the superfluid solution disappears when the diffusive gapless mode appears at $\mu_{\rm eff, R}\rightarrow -0$~\cite{PhysRevA.107.033331}.

Performing the Bogoliubov transformation, we obtain the ground-state energy 
\begin{align}
    E_{\rm GS}
        &=
     \left(\frac{m^2r\mu}{4}-\frac{m}{4a}\right)D_0^2\cr
    &+\frac{1}{2}\sum_{k}\left(\tilde{\xi}_{k}-E_{k}+mD_0^2\right),
\end{align}
where $E_{{k}}=\sqrt{\tilde{\xi}_{{k}}^2+D_0^2k^2}$
is the quasiparticle dispersion.
The value of $D_0$ can be determined from
the gap equation given by
\begin{align}
\label{eq:gapeq}
        &\frac{m^2r}{4}\left[2\mu_{\rm eff}-g_3\left( n+\frac{rm^2D_0^2}{2}\right)\right]
    -\frac{m}{2a}\cr
    & \quad\quad\quad =\sum_{k}k^2\left(\frac{1}{2E_{{k}}}-\frac{m}{k^2}\right),
\end{align}
where the number density $n$ reads
\begin{align}
    n=\frac{1}{2}
    \sum_{{k}}\left(1
    -\frac{\tilde{\xi}_{k}}{E_{{k}}}
    \right)
    -\frac{rm^2D_0^2}{4},
\end{align}
which is kept to be a real value.
Also, it is useful to define the effective complex scattering length $a_{\rm eff}$ as
\begin{align}
\label{eq:a_eff}
    \frac{1}{a_{\rm eff}}=\frac{1}{a}
    +\frac{mr}{2} \left[g_3 \left(n+\frac{rm^2D_0^2}{2}\right) -2\mu_{\rm eff} \right].
\end{align}
Using Eq.~\eqref{eq:a_eff}, one can rewrite Eq.~\eqref{eq:gapeq} as
\begin{align}
    -\frac{m}{2a_{\rm eff}}=
    \sum_{k}k^2\left(\frac{1}{2E_{{k}}}-\frac{m}{k^2}\right),
\end{align}
which is equivalent to the complex gap equation in Ref.~\cite{PhysRevA.107.033331}.
In this regard, the non-Hermitian loss effect is characterized by $1/a_{\rm eff}=1/a_{\rm eff,R}+i/a_{\rm eff,L}$ where 
\begin{align}
\label{eq:a_effI}
    \frac{1}{a_{\rm eff,I}}= 
    \frac{mr}{2}{\rm Im}\left[g_3\left( n+\frac{rm^2D_0^2}{2}\right)-2\mu_{\rm eff,I}\right],
\end{align}
indicating that the value of $rg_3$ plays a crucial role in the stability of the non-Hermitian superfluid state.
We note that $a_{\rm eff,R}\neq{\rm Re}[a_{\rm eff}]$, and $a_{\rm eff,I}\neq{\rm Im}[a_{\rm eff}]$.
Since the zero-range limit is not prohibited in the one-dimensional $p$-wave scattering~\cite{PhysRevLett.82.2536,PhysRevA.94.063650,PhysRevA.103.043307,sekino2018comparative,PhysRevA.104.023319},
our result indicates that the three-body loss process may be avoided by reducing the magnitude of the effective range, which is possible by using the optical control method~\cite{Haibin2012,Arunkmar2019,PhysRevA.106.023322}.
This is one of the main results of this paper.
It is worth noting that in one dimension the effect of the confinement-induced resonance should also be taken into account~\cite{PhysRevLett.100.170404,PhysRevLett.112.250401}.

We note that the existence of the superfluid solution does not immediately indicate its stability.
It is argued that there is a first-order-like phase transition between the normal and superfluid phases~\cite{PhysRevA.103.013724,PhysRevA.107.033331}.
To see this, one may calculate the free energy density
\begin{align}
    F_{\rm S}=E_{\rm GS}+\mu n,
\end{align}
and compare it to that in the normal phase denoted by $F_{\rm N}$.
The first-order phase transition is defined through ${\rm Re}(F_{\rm S}-F_{\rm N})=0$~\cite{PhysRevLett.123.123601,PhysRevA.103.013724,PhysRevA.107.033331,takemori2023theory}.

While we employ the mean-field theory,
it is known that such a theory works qualitatively well for describing ground-state properties in equilibrium systems throughout the BCS-BEC crossover~\cite{ohashi2020bcs}.
Although the off-diagonal long-range order is prohibited in one dimension due to strong quantum fluctuations~\cite{PhysRevLett.17.1307,PhysRev.158.383},
the mean-field theory is still useful to see characteristic features of the pairing effect in one-dimensional systems~\cite{kitaev2001unpaired}.
In this regard, our framework would be sufficient for the first examination of the $p$-wave superfluid in a one-dimensional spin-polarized Fermi gas with the three-body loss.

\subsection{Lindblad master equation for the atom-dimer inelastic collision process}
To relate $g_3$ to the observed atom-dimer inelastic collision rate $K_{\rm ad}$, we consider the Lindlad quantum master equation of the density matrix $\rho$ given by~\cite{PhysRevD.94.056006,PhysRevA.95.012708}
\begin{align}
\label{eq:lindblad}
    i\frac{d}{dt}\rho=[H,\rho]-i\{K,\rho\}
    +i\int dx L(x)\rho L^\dag(x),
\end{align}
where $H=H_{\rm a}+H_{\rm d}+V_{2}$ in the real-space basis, $L(x)$ is the Lindblad operator given by
\begin{align}
    L(x)=\sqrt{-2{\rm Im}[g_3]}\psi(x)\Psi(x) ,
\end{align}
and
\begin{align}
    K=\frac{1}{2}\int dx L^\dag(x) L(x) 
\end{align}
is the non-Hermitian part of the effective Hamiltonian.
$\psi(x)$ and $\Psi(x)$ are field operators of a Fermi atom and a Feshbach dimer, respectively, where $x$ is the one-dimensional position coordinate.
We note that the non-Hermitian effective Hamiltonian corresponds to $H_{\rm eff}=H-iK$.
While we ignored the quantum jump term (i.e., the third term of Eq.~\eqref{eq:lindblad}) in the non-Hermitian BCS-Leggett theory, we here keep it to see the connection with the observed loss rate in experiments without the postselection.

In this framework, the time-dependent atomic number $N_{\rm a}={\rm Tr}[\rho \hat{N}_{\rm a}]$ with $\hat{N}_{\rm a}=\int dx \psi^\dag(x)\psi(x)$ obeys the rate equation
\begin{align}
\label{eq:lossrate}
    \frac{d}{dt}N_{\rm a}
    &\simeq-i\langle[V_2,\hat{N}_{\rm a}]\rangle+2{\rm Im}[g_3]
    \frac{N_{\rm a}N_{\rm d}}{\ell},
\end{align}
where $N_{\rm d}={\rm Tr}[\rho \hat{N_{\rm d}}]$ with $\hat{N}_{\rm d}=\int dx \Psi^\dag(x)\Psi(x)$ is the number of Feshbach dimers and $\ell$ is the system length.
The first term with $\langle[V_2,\hat{N}_{\rm a}]\rangle\equiv {\rm Tr}[\rho[V_2,\hat{N}_{\rm a}]]$ in Eq.~\eqref{eq:lossrate} is responsible for a conversion process from atoms to dimers and a one-body decay of dimers via the Feshbach coupling
as
\begin{align}
    i\langle[V_2,\hat{N}_{\rm a}]\rangle\simeq 
    -2\Gamma N_{\rm d} +2K_{\rm aa}\frac{N_{\rm a}^2}{2\ell},
\end{align}
where $\Gamma=-\ell{\rm Im}\Sigma_{\rm d}^{\rm ret.}(P\simeq \bar{P}_{\rm d},\omega\simeq \bar{\omega}_{\rm d})$
and $K_{\rm aa}=-2\ell {\rm Im}\Sigma_{\Pi}^{\rm ret.}(P\simeq \bar{P}_{\rm d},k\simeq \bar{k}_{\rm a},\omega\simeq \bar{\omega}_{\Pi})$ are the dimer and two-particle retarded self-energies at the averaged dimer and atom momenta $\bar{P}_{\rm d}$ and $\bar{k}_{\rm a}$, and dimer and two-particle frequencies $\bar{\omega}_{\rm d}$ and $\bar{\omega}_{\Pi}$, respectively 
(see Appendix \ref{app:a} where the relation to the two-step cascade model is microscopically discussed within the Schwinger-Keldysh approach~\cite{rammer2011quantum}). 
Here we focus on the second term of Eq.~\eqref{eq:lossrate} and for simplicity, we consider the homogeneous system where the number densities are given by $n_{\rm a,d}=N_{\rm a,d}/\ell$ with the system length $\ell$ (which is approximately given by the twice Thomas-Fermi radius as $\ell \simeq 2R_{\rm R}$ for comparison with Ref.~\cite{PhysRevLett.125.263402}).
In this regard, one finds 
\begin{align}
    K_{\rm ad}=-2{\rm Im}[g_3].
\end{align}

Using the experimental values reported in Ref.~\cite{PhysRevLett.125.263402}, as the atom-dimer collision rate $K_{\rm ad}\simeq 0.67 \ {\rm cm/s}$, the system length $\ell \simeq 2R_{\rm F}\simeq 2\times 10^{-3} \ {\rm cm}$, the total number $N_{\rm a}\simeq 9\times 10^4$,
and the Fermi energy $E_{\rm F}\simeq 4.8 \ \mu{\rm K}$,
we can estimate the normalized atom-dimer coupling 
$\gamma_3=-{\rm Im}[g_3]\frac{n_{\rm a}}{E_{\rm F}}$ as
$\gamma_3\simeq 1.5\times 10^{2}$, which is much larger than the typical many-body energy scale (i.e., Fermi energy and momentum).
Incidentally, under the assumption of the dimer steady state, the three-body loss rate is given by $L_3=\frac{3}{2}\frac{K_{\rm aa}}{\Gamma}K_{\rm ad}$~\cite{PhysRevLett.125.263402}.
The accurate calculations of $K_{\rm aa}$ and $\Gamma$ will be left for future work.

\section{Results}
\label{sec:3}

\begin{figure}[t]
    \centering
    \includegraphics[width=8.6cm]{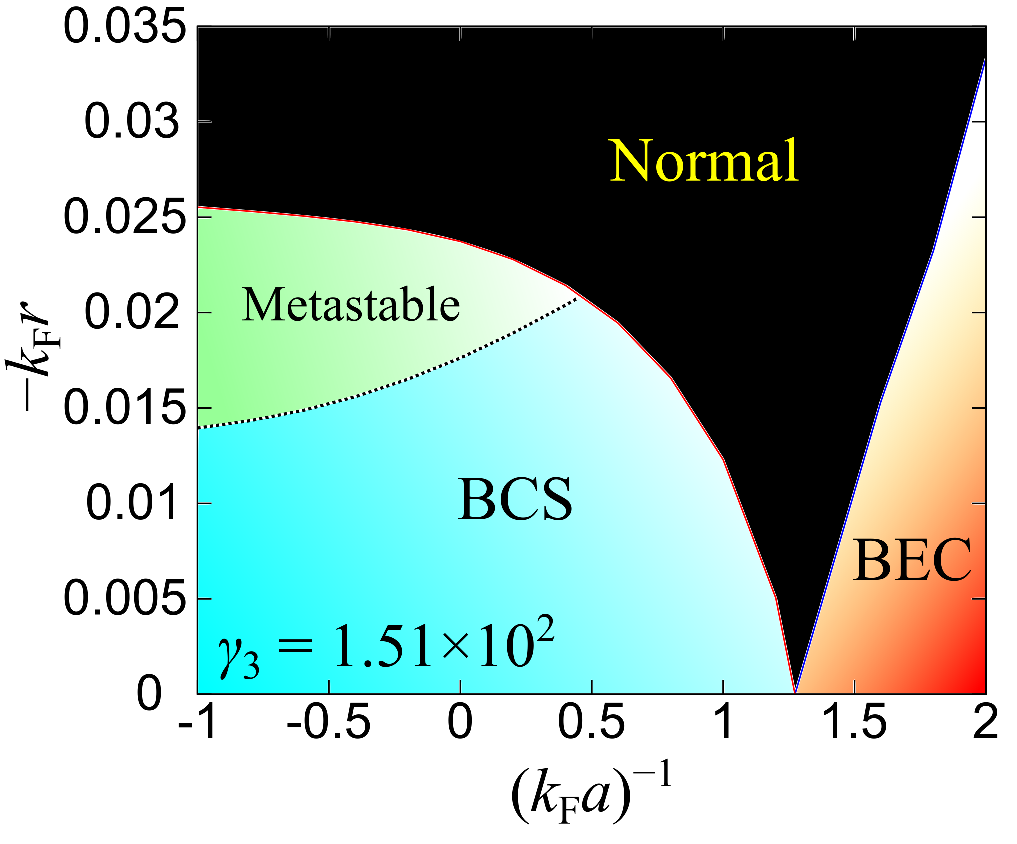}
    \caption{Calculated ground-state phase diagram with the normalized imaginary three-body coupling $\gamma_3=1.5\times 10^{2}$.
    The BCS state becomes metastable when the BCS free energy $F$ is larger than that of the normal phase.
    The dotted curve represents the first-order-like phase transition between the BCS and the normal phase.
    The boundary between the BCS phase (or metastable BCS phase) and the normal phase and that between the BEC phase and the normal phase are accompanied by the appearance of the exceptional point and the diffusive gapless mode, respectively~\cite{PhysRevA.107.033331}.
    }
    \label{fig:1}
\end{figure}
Figure~\ref{fig:1} shows the mean-field ground-state phase diagram of the present non-Hermitian system in the plane of the inverse $p$-wave scattering length $(k_{\rm F}a)^{-1}$ and effective range $-k_{\rm F}r$ normalized by the Fermi momentum $k_{\rm F}$,
where the normalized three-body loss parameter is taken as $\gamma_3=1.5\times 10^2$ that is relevant to the experiment~\cite{PhysRevLett.125.263402}.
In the numerical calculation, we used a sufficiently large cutoff as $\Lambda/k_{\rm F}=10^2$ where we confirmed that the numerical results are qualitatively unchanged.
One may find that the structure of the phase diagram is similar to that in two-component Fermi gases with two-body losses~\cite{PhysRevA.107.033331}.
For the zero-range limit ($-k_{\rm F}r\rightarrow 0$) where $\mu_{\rm eff}$ is equal to $\mu$ because of the vanishing dimer fraction (i.e., $\phi_0^2\rightarrow 0$),
the system exhibits the topological phase transition from the BCS phase [${\rm Re}(\mu)>0$] to the BEC phase [${\rm Re}(\mu)<0$] at $(k_{\rm F}a)^{-1}=4/\pi\simeq 1.27$~\cite{PhysRevB.105.064508}.
Around this point,
the system is extremely fragile against the non-Hermitian loss term as found in the case with the two-body loss~\cite{PhysRevA.107.033331}.
However, since the three-body loss occurs via the atom-dimer collision,
the present system is exceptionally stable against the three-body loss effect in the zero-range limit with a vanishing dimer fraction.
If one increases $k_{\rm F}|r|$,
the loss effect becomes remarkable as the superfluid solution disappears at the exceptional point (i.e., $z\rightarrow 0$) in the BCS phase and at the appearance of the diffusive gapless mode in the BEC phase (i.e., $\mu_{\rm eff, R}\rightarrow -0$).
Moreover, the weak-coupling BCS phase can be metastable against the normal phase due to the atomic loss effect as it is found in three-dimensional $s$-wave supefluid~\cite{PhysRevA.103.013724} as well as in one-dimensional two-component $p$-wave superfluid.~\cite{PhysRevA.107.033331}. 

In the experiment~\cite{PhysRevLett.125.263402},
the quasi-one-dimensional effective range is given by $r=R_p a_\perp^2/6$
where $R_p=-0.41(1)a_0^{-1}$ is the $p$-wave effective range ($a_0$ is the Bohr radius) in three dimensions and $a_\perp=\sqrt{2/m\omega_r}$ is the transverse oscillator length with the frequency $\omega_r$.
Using the experimental value $\omega_{r}=\sqrt{4\times 75}E_{r}$ where $E_r=1.41 \ \mu {\rm K}$ is the recoil energy, 
one can estimate $r\simeq -4.3 \ \mu {\rm m}$ and hence $k_{\rm F}r\simeq -47$.
This value is much far away from the superfluid region.

The relation with the two-component system involving two-body loss~\cite{PhysRevA.107.033331} is more evident if one examines the imaginary part of the effective scattering length. 
Because of extremely large $\gamma_3$,
one may approximately obtain the inverse effective scattering length
\begin{align}
\label{eq:aI}
    \frac{1}{k_{\rm F}a_{\rm eff,I}}\simeq  \frac{1}{4}\gamma_3k_{\rm F}|r|,
\end{align}
indicating $(k_{\rm F}a_{\rm eff,I})^{-1}\simeq 0.9$ around $(k_{\rm F}a)^{-1}=-1$, which is consistent with Ref.~\cite{PhysRevA.107.033331}
when replacing $a_{\rm eff, I}$ with the imaginary scattering length induced by the two-body loss.

\begin{figure}[t]
    \centering
    \includegraphics[width=8.6cm]{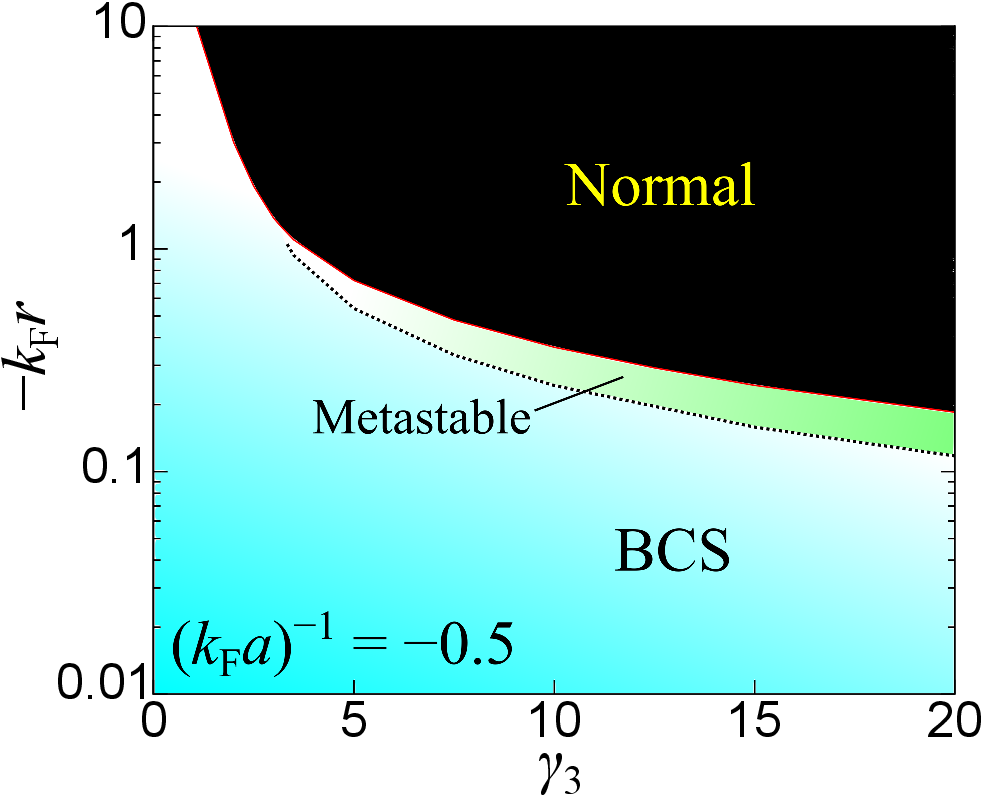}
    \caption{Calculated ground-state phase diagram in the plane of the normalized imaginary three-body coupling $\gamma_3$ and the normalized effective range $-k_{\rm F}r$. In this figure, $(k_{\rm F}a)^{-1}=-0.5$ is used. }
    \label{fig:2}
\end{figure}

To examine the dependence on $\gamma_3$,
we address the ground-state phase diagram with respect to $\gamma_3$ and $-k_{\rm F}r$ at $(k_{\rm F}a)^{-1}=-0.5$ as shown in Fig.~\ref{fig:2}.
For smaller $\gamma_3$, the superfluid solution is allowed up to larger $k_{\rm F}|r|$.
The magnitude of the critical effective range of the superfluid solution becomes smaller and smaller with increasing $\gamma_3$.
As shown in Eq.~\eqref{eq:aI},
for larger $\gamma_3$ the critical effective range is proportional to $1/\gamma_3$.
We note that
the first-order-like phase transition is absent in the case with a large negative effective range.
This can be understood as reaching the strong-coupling side induced by the negative range effect~\cite{PhysRevLett.108.250401,PhysRevA.97.043613,tajima2019generalized,PhysRevA.101.013615},
where the metastable superfluid solution exists in only the weak-couplig regime~\cite{PhysRevA.107.033331}.

While we consider a pure imaginary atom-dimer coupling ${\rm Im}[g_3]$ to examine the three-body loss process,
the real part ${\rm Re}[g_3]$ can also be present as pointed out in Ref.~\cite{PhysRevA.103.043307}.
In such a case, $\mu_{\rm eff}$ is modified by $-{\rm Re}[g_3]\phi_0^2$, so that the exceptional point and the apearance of the difussive gapless mode would be shifted quantitatively. 
However, it is out of scope in this work.

\section{Summary}
\label{sec:4}
To summarize,
in this work, we have discussed the non-Hermitian $p$-wave Fermi superfluid near the $p$-wave Feshbach resonance involving inelastic atom-dimer collisions, which lead to the three-body recombination via the two-step cascade process.
Developing the non-Hermitian BCS-Leggett theory for the present system, we have shown how the superfluid state is destroyed by the inelastic atom-dimer collision.
Importantly, the existence of the superfluid solution is deeply related to the magnitude of the $p$-wave effective range $r$, which characterizes the coupling between Fermionic atoms and Feshbach dimers. 
It is found that the smaller value of $k_{\rm F}|r|$
is highly advantageous for the realizing non-Hermitian $p$-wave Fermi superfluid state under the postselected quantum trajectories.
Estimating the imaginary atom-dimer coupling from the experimental result~\cite{PhysRevLett.125.263402},
we have drawn the possible ground-state phase diagram in the plane of the $p$-wave low-energy constants.
Our result suggests that the $p$-wave Fermi superfluid state can be achieved by reducing $k_{\rm F}|r|$ with e.g., the optical control of scattering properties~\cite{Haibin2012,Arunkmar2019,PhysRevA.106.023322} and the strong transverse trap confinement~\cite{PhysRevLett.100.170404,PhysRevLett.112.250401}, where the one-dimensional zero-range limit is not prohibited by the causality bound~\cite{hammer2009causality,hammer2010causality} in contrast to higher dimensions.

For future perspectives,
it is interesting to consider the beyond-mean-field effects by including pairing fluctuations.
The inclusion of the jump term and the self-consistent calculation of $K_{\rm aa}$ and $\Gamma$ during the time evolution would enable us to perform a more detailed comparison with the experiment.
The effects of the confining trap potential and quasi-one-dimensionality should also be addressed for further quantitative investigations.
Moreover, our formalism for the three-body loss can be applied to studies of the non-Hermitian Efimov effect~\cite{PhysRevResearch.3.043225,PhysRevResearch.5.043010}.

\acknowledgements
H.~T. was supported by the JSPS Grants-in-Aid for Scientific Research under Grants No.~18H05406, No.~22H01158, and No.~22K13981. A.~D. was supported by JSPS KAKENHI Grant No. 23K19056.
T.~H. was supported by the JSPS Grants-in-Aid for Scientific Research under Grants No.~21H01007, and No.~21H01084.
Y.~S. and S.~N. were supported by Pioneering Program of RIKEN for Evolution of Matter in the Universe (r-EMU).
Y.~S. was supported by JST ERATO Grant Number JPMJER2302, Japan.

\appendix
\section{Two-body transition process and the relation to the two-step cascade model}
\label{app:a}
\begin{widetext}
In this appendix,
we discuss the relation with the two-step cascade model~\cite{PhysRevA.99.052704,PhysRevLett.120.193402,PhysRevLett.125.263402} by examining the two-body term $-i\langle [V_2,\hat{N}_a]\rangle$ in Eq.~\eqref{eq:lossrate}.
We need to evaluate
\begin{align}
        i\langle[V_2,\hat{N}_{\rm a}]\rangle
    &=2\sum_{P,k}
    kg_2{\rm Re}[i\langle d_P^\dag a_{-k+P/2}a_{k+P/2}\rangle].
\end{align}
The expectation value $\langle \cdots\rangle$ should be a two-time correlation function for an infinitesimally small time step $dt$ as $\langle \cdots\rangle=\langle \Phi(t')|\cdots|\Phi(t)\rangle$ with $t'=t+dt$ where $|\Phi(t)\rangle$ describes the time-dependent many-body state vector.
While we are interested in the strongly-interacting regime,
the lowest-order perturbation with respect to the atom-dimer transition $V_2$ can be a good approximation since 
the transition rate during the infinitesimally short time period $dt$ is concerned here. 
Accordingly, we consider the perturbation on the Schwinger-Keldysh contour for the two-time expectation value~\cite{rammer2011quantum} as
\begin{align}
    \langle d_P^\dag(t') a_{-k+P/2}(t)a_{k+P/2}(t)\rangle
    &=\langle 
    T_{C}\left[
    e^{-i\int_C dt''V_2(t'')}
    d_P^\dag(t') a_{-k+P/2}(t)a_{k+P/2}(t)
    \right]\rangle\cr
    &=-i\int_C dt''\langle T_C[V_2(t'') d_P^\dag(t') a_{-k+P/2}(t)a_{k+P/2}(t)]\rangle +O(V_2^2),
\end{align}
where $T_C$ denotes the time-order product on the deformed time contour $C$, and $d_P(t)$, $a_{k}(t)$ and $V_2(t)$ are the Heisenberg representations of the operators.
Considering the Wick theoreom, one can find the relevant contribution given by
\begin{align}
    \langle d_P^\dag(t') a_{-k+P/2}(t)a_{k+P/2}(t)\rangle
    &=2i\int_C dt''
    kg_2
    G_{\rm d}(P,t'',t')
    \Pi(P,k,t,t''),
    \label{eq:a3}
\end{align}
where
\begin{align}
iG_{\rm d}(P,t'',t')=    \langle T_C[d_P(t'')d_P^\dag(t')]
    \rangle,
    \end{align}
and
\begin{align}
    i\Pi(P,k,t,t'')
    =
    \langle T_C[a_{-k+P/2}(t)a_{k+P/2}(t)a_{k+P/2}^\dag(t'')
    a_{-k+P/2}^\dag (t'')]\rangle,
\end{align}
are the contour-ordered Green's functions of a Feshbach dimer and two Fermi atoms, respectively.
Since our aim is to derive the two-step cascade process used in the experimental analysis of normal-state $p$-wave Fermi gases~\cite{PhysRevLett.125.263402}, the condensates and the anomalous Green's functions are ignored here.
Using the Langreth rule~\cite{rammer2011quantum},
we further rewrite Eq.~\eqref{eq:a3} as
\begin{align}
    \langle d_P^\dag(t') a_{-k+P/2}(t)a_{k+P/2}(t)\rangle
    &=-2i\int_{-\infty}^{\infty}dt''
    kg_2
    [\Pi^{\rm ret.}(P,k,t,t'')G_{\rm d}^{<}(P,t'',t')
    +\Pi^{<}(P,k,t,t'')G_{\rm d}^{\rm adv.}(P,t'',t')],
\end{align}
Here we introduced the lesser Green's functions
\begin{align}
    iG_{\rm d}^{<}(P,t'',t')
    =\langle d_P^\dag (t') d_P(t'')\rangle, 
\end{align}
\begin{align}
    i\Pi^{<}(P,k,t,t'')
    =
    \langle
    a_{k+P/2}^\dag(t'')
    a_{-k+P/2}^\dag (t'')
    a_{-k+P/2}(t)
    a_{k+P/2}(t)\rangle,
\end{align}
the retarded two-particle Green's function
\begin{align}
    i\Pi^{\rm ret.}(P,k,t,t'')
    =\theta(t-t'')\langle[a_{-k+P/2}(t)a_{k+P/2}(t),a_{k+P/2}^\dag(t'') a_{-k+P/2}^\dag(t'')]\rangle,
\end{align}
and the advanced dimer Green's function
\begin{align}
    iG_{\rm d}^{\rm adv.}(P,t'',t')=\theta(t-t'')
    \langle[d_P(t''),d_P^\dag(t')]\rangle.
\end{align}
We note $G_{\rm d}^{\rm adv.}(P,t'',t')=[G_{\rm d}^{\rm ret.}(P,t',t'')]^*$.

To make further progress, we assume the non-equlibrium steady state with the time-translational symmetry
which allows us to perform the Fourier transformation as
\begin{align}
       i\langle 
       [V_2,\hat{N}_{\rm a}]
       \rangle
&=4\sum_{P}\int_{-\infty}^{\infty}\frac{d\omega}{2\pi}
    {\rm Re}\left[\Sigma_{\rm d}^{\rm ret.}(P,\omega)G_{\rm d}^{<}(P,\omega)
    +
    \sum_{k}
    \Pi^{<}(P,k,\omega)[\Sigma_{\Pi}^{\rm ret.}(P,k,\omega)]^*\right],
\end{align}
where we introduced the dimer retarded self-energy 
\begin{align}
    \Sigma_{\rm d}^{\rm ret.}(P,\omega)
    =g_2^2\sum_{k}k^2\Pi^{\rm ret.}(P,k,\omega),
\end{align}
and two-particle retarded self-energy
\begin{align}
    \Sigma_{\Pi}^{\rm ret.}(P,k,\omega)
    =g_2^2k^2G_{\rm d}^{\rm ret.}(P,\omega).
\end{align}
\begin{figure}
    \centering
    \includegraphics[width=5cm]{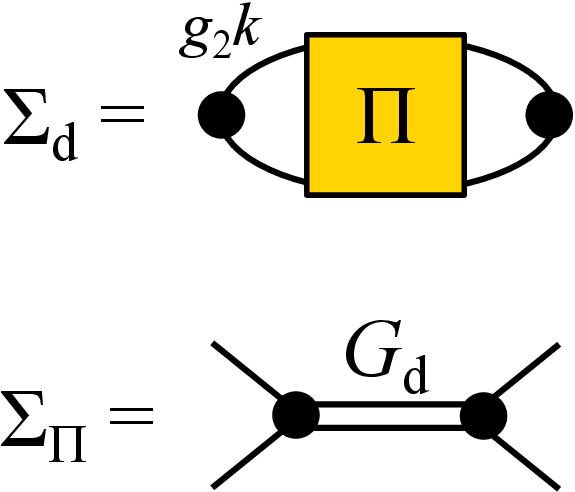}
    \caption{Diagramatic representation of the dimer self-energy $\Sigma_{\rm d}$ and the two-particle self-energy $\Sigma_{\Pi}$. The black dot denotes the $p$-wave Feshbach coupling $g_2k$. The box and double solid lines represent the two-particle propagator $\Pi$ and the dimer propagator $G_{\rm d}$, respectively.}
    \label{fig:3}
\end{figure}
As diagrammatically shown in Fig.~\ref{fig:3},
while $\Sigma_{\rm d}^{\rm ret.}(P,\omega)$ describes the decay of a dimer into two atoms and its inverse process,
$\Sigma_{\Pi}^{\rm ret.}(P,k,\omega)$ does the formation of a dimer and its inverse process.
The lesser components can be rewritten approximately by the equilibrium values as
$iG_{\rm d}^{<}(P,\omega)\simeq  b(\omega)A_{\rm d}(P,\omega)$ and
$   i\Pi^{<}(P,k,\omega)
    \simeq b(\omega)A_{\Pi}(P,k,\omega)$
where $A_{\rm d}(P,\omega)$ and $A_{\Pi}(P,k,\omega)$ are dimer and two-particle spectral functions, respectively.
$b(\omega)$ is the Bose-Einstein distribution funciton.
Using them,
we obtain
\begin{align}
       i\langle 
       [V_2,\hat{N}_{\rm a}]
       \rangle
&=4\sum_{P}\int_{-\infty}^{\infty}\frac{d\omega}{2\pi}
   \left[
    b(\omega)A_{\rm d}(P,\omega)
    {\rm Im}\Sigma_{\rm d}^{\rm ret.}(P,\omega)
    -
    \sum_{k}
    b(\omega)A_{\Pi}(P,k,\omega)
    {\rm Im}\Sigma_{\Pi}^{\rm ret.}(P,k,\omega)\right] .
\end{align}
Motivated by the experimental work~\cite{PhysRevLett.125.263402},
we can define the lifetime of the dressed dimer 
\begin{align}
\Gamma=-2\ell{\rm Im}\Sigma_{\rm d}^{\rm ret.}(P\simeq \bar{P}_{\rm d},\omega\simeq\bar{\omega}_{\rm d}) ,
\end{align}
and the two-body inelastic collision rate 
\begin{align}
    K_{\rm aa}=-4\ell {\rm Im}\Sigma_{\Pi}^{\rm ret.}(P\simeq \bar{P}_{\rm d},k\simeq \bar{k}_{\rm a},\omega\simeq \bar{\omega}_{\Pi}),
\end{align}
where we ignored the momentum- and frequency-dependence of the self-energies by replacing them with the averaged values $\bar{P}_{\rm d}=\sum_{P}|P|f_{{\rm d},P}$, $k_{\rm a}=\sum_{k}|k|f_{{\rm d},k}$, $\bar{\omega}_{{\rm d}}=\sum_{P}\frac{P^2}{4m}f_{{\rm d},P}$, and $\bar{\omega}_{\Pi}=\sum_{P,k}\left[\frac{(P+k/2)^2}{2m}f_{{\rm a},P+k/2}+\frac{(P-k/2)^2}{2m}f_{{\rm a},P-k/2}\right]$  (noting that $f_{{\rm d},P}$ and $f_{{\rm a},k}$ are the momentum distributions of dimers and atoms, respectively).
Eventually, the rate equation of $N_{\rm a}$ is given by
\begin{align}
\label{eq:cascade}
    \frac{dN_{\rm a}}{dt}
    \simeq 
    2\Gamma N_{\rm d}
    -2K_{\rm aa}\frac{N_{\rm a}^2}{2\ell}
    -K_{\rm ad}\frac{N_{\rm a}N_{\rm d}}{\ell},
\end{align}
where we used
$\sum_{P}\int\frac{d\omega}{2\pi}b(\omega)A_{\rm d}(P,\omega)=\frac{N_{\rm d}}{\ell}$
and
$\sum_{P,k}
    \int\frac{d\omega}{2\pi}b(\omega)A_{\Pi}(P,k,\omega)
    \equiv i\Pi^{<}(P,k,\omega)
    \simeq
    \sum_{P,k} f_{{\rm a},P+k/2} f_{{\rm a},-P+k/2}
    =\frac{N_{\rm a}^2}{\ell^2}
$
 assuming the homogeneous system with the system length $\ell$ and ignoring the atomic self-energy.
Equation~\eqref{eq:cascade} is indeed consistent with the two-step cascade model considered in Ref.~\cite{PhysRevLett.125.263402} except for the second term of Eq.~\eqref{eq:cascade}  because we consider the large-$N_{\rm a}$ limit (i.e., $N_{\rm a}\rightarrow \infty$) whereas it is proportional to $N_{\rm a}(N_{\rm a}-1)$ in Ref.~\cite{PhysRevLett.125.263402}.
The loss rate equation of the dimer number can also be obtained similarly.

In this way, one can see the relation between our model and the two-step cascade model.
It is reported that the three-body loss rate $L_3$ is proportional to $K_{\rm aa}K_{\rm ad}/\Gamma$ by assuming the steady state of dimers.
While $K_{\rm aa}$ is proportional to $g^2$ and hence $r^{-1}$ at the leading order,
this $r^{-1}$ behavior is compensated by the denominator $\Gamma$ which is also propotional to $r^{-1}$.
Thus, $L_3$ is independent of $r$ at this approximation, and our conclusion that a smaller magnitude of the effective range is more advantageous for realizing the $p$-wave superfluid state is unchanged. 
\end{widetext}

\bibliographystyle{apsrev4-2}
\bibliography{reference.bib}

\end{document}